
\input phyzzx

\hfuzz=10pt

\def\ie{{\it i.e.}}

\def\IJMP{{\it Int.\ Jour.\ Mod.\ Phys.\ }}

\def\MPL{{\it Mod.\ Phys.\ Lett.\ }}

\def\NP{{\it Nucl.\ Phys.\ }}
\def\PL{{\it Phys.\ Lett.\ }}
\def\PR{{\it Phys.\ Rev.\ }}

\def\PRL{{\it Phys.\ Rev.\ Lett.\ }}

\def\refmark#1{[#1]}		

\def\Exp#1{{\rm e}^{#1}}

\pubtype={}

\def\vev#1{\langle #1 \rangle}

\def\square{\hbox{{$\sqcup$}\llap{$\sqcap$}}}   

\font\eightrm=cmr8
\font\eightit=cmti8

\overfullrule=0pt

\def\refmark#1{[#1]}		
\def\refmark#1{$^{[#1]}$}
\def\art#1{}
\def\art#1{{\sl #1}}

\def\Chapter#1#2{\bigskip\noindent{\tenbf {\tenbf #1\ \ #2 \hfill}}}
\def\Section#1#2{\smallskip\noindent{\tenrm #1 {\tenit\ \ #2 \hfill}}}
\def\OMIT#1{}

\def\vev#1{\langle #1 \rangle}
\def\square{\hbox{{$\sqcup$}\llap{$\sqcap$}}}   

\def\phi{\varphi}
\def\ap{\alpha'}
\def\d{{\rm d}}
\def\D{{\rm D}}
\def\rdot{{\dot r}}
\def\xdot{{\dot x}}
\def\xddot{{\ddot x}}
\def\xidot{{\dot \xi}}
\def\xiddot{{\ddot \xi}}
\def\etadot{{\dot \eta}}
\def\etaddot{{\ddot \eta}}
\def\rdot{{\dot r}}
\def\phidot{{\dot \phi}}
\def\lpl{{\ell_{\rm Planck}}}
\def\Bdot{{\dot B_0}}
\def\Bddot{{\ddot B_0}}

\def\PRS{{\it Proc.\ Roy.\ Soc.\ }}

\REF\turok{
   N.~Turok and P.~Bhattacharjee,
   \art{Stretching cosmic strings,}
   \PR {\bf D29} (1984) 1557;
   N.~Turok,
   \art{String-driven inflation,}
   \PRL {\bf 60} (1988) 549
}

\REF\shockwave{D.~Amati and C.~Klimcik,
   \art{Strings in a shock wave background and generation of curved
   geometry from flat-space string theory,}
   \PL {\bf 210B} (1988) 92
}

\REF\horowitz{
   G.~T.~Horowitz and A.~R.~Steif,
   \art{Spacetime singularities in string theory,}
   \PRL {\bf 64} (1990) 260;
   \art{Strings in strong gravitational fields,}
   \PR {\bf D42} (1990) 1950;
   G.T.~Horowitz,
   \art{Singularities in string theory,}
   UCSB-TH-90-34, 1990;
}

\REF\devegaB{
   H.J.~de Vega and N.~S\'anchez,
   \art{Mass and energy-momentum tensor of quantum strings in gravitational
   shock waves,}
   \IJMP {\bf A7} (1992) 3043;
   \art{Strings falling into space-time singularities,}
   \PR {\bf D45} (1992) 2783
}

\REF\devega{
   H.~J.~de Vega and N.~S\'anchez,
   \art{A new approach to string quantization in curved spacetimes,}
   \PL {\bf 197B} 320;
   \art{Quantum dynamics of strings in black hole spacetimes,}
   \NP {\bf B309} (1988) 552;
   \art{The scattering of strings by a black hole,}
   \NP {\bf B309} (1988) 577
}

\REF\minlength{
   D.J.~Gross,
   \art{Strings at superplanckian energies: in search of the string symmetry,}
   {\it Phil.\ Trans.\ R.\ Soc.\ Lond.\ } {\bf A329} (1989) 401;
   D.~Amati, M.~Ciafaloni, and G.~Veneziano,
   \art{Can spacetime be probed below the string size?}
   \PL {\bf 216B} (1989) 41;
   G.~Veneziano,
   \art{An enlarged uncertainty principle from gedanken string collisions?}
   CERN-TH.5366/89, 1989;
   \art{Quantum string gravity near the Planck scale,}
   CERN-TH.5889/90, 1990;
   K. Konishi, G.~Paffuti and P~.Provero,
   \art{Minimum physical length and the generalized uncertainty principle
   in string theory,}
   \PL {\bf B234} (1990) 276;
   R.~Guida, K.~Konishi and P.~Provero,
   \art{On the short distance behavior of string theories,}
   \MPL {\bf A6} (1991) 1487;
}

\REF\highsym{D.J.~Gross,
   \art{High energy symmetries of string theory,}
   \PRL {\bf 60} (1988) 1229 }

\REF\GMa{D.J.~Gross and P.F.~Mende,
   \art{High energy behavior of string scattering amplitudes,}
   \PL {\bf 197B} (1987) 129;
   \art{String theory beyond the Planck scale,}
   \NP {\bf B303} (1988) 407}

\REF\MO{P.F.~Mende and H.~Ooguri,
   \art{Borel summation of string theory for Planck scale
   scattering,}
   \NP {\bf B339} (1990) 641 }

\REF\papapetrou{A.~Papapetrou,
   \art{Spinning test-particles in general relativity I,}
   \PRS {\bf A209} (1951) 248;
   E.~Corinaldesi and A.~Papapetrou,
   \art{Spinning test-particles in general relativity II,}
   \PRS {\bf A209} (1951) 259;
}

\date={July 1992}
\Pubnum{Brown-HET-875}
\pubtype={hep-th/9210001}

\titlepage
\hoffset=0.5truein
\voffset=0.5truein
\hsize=5truein
\vsize=7.8truein
\baselineskip=13pt
\tenpoint

\vglue.5in

\titlestyle{{\tenbf STRING THEORY AT SHORT DISTANCE AND \break
THE PRINCIPLE OF EQUIVALENCE}
\footnote{\star}{Lecture delivered at Workshop on String Quantum Gravity and
Physics at the Planck Energy Scale, Erice, June 1992.
E-mail: {\tt mende@het.brown.edu }
}
}

\bigskip

\titlestyle{\tenrm PAUL F.~MENDE
}

\titlestyle{
\eightit
   Department of Physics,  Brown University  \break
   Providence, Rhode Island~~02912, USA
   }

\bigskip

\titlestyle{\tenrm ABSTRACT}
{
\eightrm
\baselineskip=10pt
\narrower \parindent=1.5pc
Point particles fall freely along geodesics; strings do not.
In string theory all probes of spacetime structure, including
photons, are extended objects and therefore always subject
to tidal forces.  We illustrate how string theory modifies
the behavior of light in weak gravitational fields and limits
the applicability of the principle of equivalence.
This gives in principle a window on the short-distance structure
of geometry in quantum gravity where one can see
in a model-independent way how some of
its predictions differ from those of classical gravity.
We compare this with the lessons of high-energy string scattering.
\smallskip}


\tenrm
\baselineskip=13pt

\Chapter{1.}{Introduction}

\noindent
In this talk I will discuss how one may get some insight into
gravity and geometry at Planckian distances.
String theory is a theory of quantum gravity, but we are far
from being able to test it and have difficulty in facing some of
the most basic issues.
We will need a better understanding of the geometric structures
if we to are fully uncover the symmetries and vacuum of the string.
By the same token, clarifying how the coupling to geometry at these distances
departs from general relativity may make it possible to identify
low-energy effects which carry the signature of string physics.
We examine how the extended structure of strings --- crucial in
cutting off the divergences of quantum gravity --- leads to interesting
qualitative departures from from classical gravity as well as from
local quantum field theory.

First we review why high-energy scattering of extended objects
 --- as opposed to point particles ---
appears incapable of probing short-distance geometry.
We then turn to low-energy manifestations of Planck scale physics
by considering propagation of test strings in background gravitational
fields.
We attempt to highlight the most fundamental difference between particles
and strings:  particles fall freely while strings are subject to
tidal forces.
These tidal forces produce qualitatively new effects that are
independent of most details of the particular string theory and
its vacuum.
\footnote{{\rm a}}{\eightrm\baselineskip=10pt
It should be noted, at least in passing, that this discussion does not
apply to the fascinating two-dimensional string theories
that have been the focus of much recent attention.
Without transverse dimensions, strings are as much like particles as they
can be, and the effects of interest will have no counterpart unless it can
be shown that some vestige of the transverse dimensions has
tidal couplings to background fields.
}
We illustrate, therefore, using the open bosonic string with little
loss of generality.
Moreover, it is not necessary to have an exact solution to
the string equations of motion for the metric for these effects
to be manifest.
Therefore these issues may be largely decoupled
from the problem of finding conformally invariant string
vacuua.

The importance of tidal forces on string propagation has been emphasized
in the context of cosmic strings\refmark{\turok}
and
in connection with the issue of singularities in string theory
\refmark{\shockwave, \horowitz, \devegaB}.
Since modes of a test string may couple to background fields that
are never seen
by a test particle, the criteria for singular propagation and geodesic
completeness are different for strings and particles.
This was studied in the case of shock-wave type
backgrounds that have the
twin virtues of being exact solutions to the string equations of motion
and of being exactly solvable, with linear equations for the string modes.

The latter `virtue,' however, is peculiar to this highly restricted
class of geometries.
For generic backgrounds, the non-linear interactions mix string modes.
We isolate the effect of this mixing
on the string zero-modes and show
that tidal forces lead to a qualitative change in photon trajectories.

\Chapter{2.}{Short Distance and High Energy}

\noindent
The traditional route to fundamental physics of elementary particles
is to explore short distances through high energies.
Forces become simpler, compositeness is revealed, symmetries are restored.
For point particles --- and the local field theories which describe them ---
short distance and high energy are almost synonymous.
For strings they are not.
High energy particles have short wavelengths,
and one may attempt to probe short distances
$\Delta x$ with high enough energy $E$ such that
$$
   \Delta x \sim 1/E
   .\eqn\?
$$
Strings have an obstruction to resolving short distances:
a string given extremely high energy can stretch to large size,
so that there may be an effective minimal length that
can be probed\refmark{\minlength},
$$
   \Delta x\, \gsim \, c_1/E + c_2 E
   ,\eqn\stringUP
$$
where $c_i$ are constants.

Recall how this comes about and how it is modified by the
large quantum fluctuations of the string.
The amplitude at each order is given by a path integral, or sum
over string histories:
$$
   {\cal A}_G(p_i) = \int {\cal D}g {\cal D}X\,  \Exp{-I[X,g] +i\int P\cdot X}
   \eqn\pathsum
$$
where $X^\mu(\xi)$ is the spacetime trajectory of the string;
$G$ is the genus of the surface;
the string action is its invariant area,
$$
   I[X,g] = {-{1 \over 2\pi} \int {\d^2\xi}
   \sqrt{g}g^{ab} \partial_{\alpha}X^{\mu}\partial_{\beta}X_{\mu} };
   \eqn\?
$$
$g_{ab}$ is the metric on the punctured worldsheet;
and $P$ is a source representing the incoming particles for
the scattering of momentum eigenstates,
$ P^\mu (\xi) = \sum p^\mu_i \delta^{(2)}(\xi,\xi_i). $

When {\it all\/} of the momenta get large (\ie, fixed-angle scattering),
the path integral is dominated by a classical trajectory and
we can evaluate it semiclassically.
This is easy to see:
define rescaled string coordinates and momenta by
$$
   X = \sqrt {s} {\tilde X},
   \qquad
   p_i = \sqrt{s} {\tilde p}_i
   ,\eqn\Xscaled
$$
where $s=4E^2$.
Then
$$
   {\cal A}_G(s;{\tilde p}_i) = \int {\cal D}g {\cal D}{\tilde X}
   \,\Exp{~s\left(-I[{\tilde X},g]
   +i\int {\tilde P}\cdot {\tilde X}\right) }.
   \eqn\?
$$
Thus the $s\to\infty$, or equivalently
$M_{\sl Planck}\to 0$, limit corresponds to
the semiclassical limit of string theory\refmark{\highsym},
with the dominant contribution coming from the trajectory
$$
   {\tilde X}^\mu(\xi)
   = \sum_i {\tilde p}^\mu_i \, {\cal G}(\xi,\xi_i)
   \eqn\Xclass
$$
where ${\cal G}(\xi,\xi')$ is the Green function
for the action $I[{\tilde X},g]$.
In the coordinates of Refs.~[\GMa],
this can be written as
$$
   X^{\mu}(\xi)= {\sqrt{s}\over G+1}i\sum_{j=1}^4
   \tilde p_i^{\mu}\ln|\xi- a_j| +
   O(1/s)\,\,.
   \eqn\surface
$$
Observe from Eq.~$\Xscaled$ that the mean size
of the string histories contributing
to the integral {\it grows with energy as ${\sqrt s}$\/}.
This observation leads to uncertainty relation of Eq.~$\stringUP$.

Using the stationary point Eq.~$\Xclass$
to evaluate the amplitude, one finds
$$
   {\cal A}_G \sim {\rm e}^{-s/(G+1)}
   .\eqn\?
$$
This behavior is extremely soft at high energies, so soft in fact
that it violates bounds for local quantum field theories.
The source of the behavior is not hard to find:  it is characteristic
of an extended object, and it is precisely the extended structure
which enters at high energies that is responsible for the finite
ultraviolet behavior that is the {\it raison d'\^etre\/} of string
theory as a theory of quantum gravity.
Certainly we need to understand better how the theory can be non-local
without spoiling causality, but can we deduce directly from these results
any information about how string theory modifies spacetime structure
at the Planck scale?

Because the high-energy limit corresponds to the semi-classical
limit, it is tempting to try to interpret the stationary trajectories
as ``the'' path of the string, and the size of the worldsheet as
``the'' size of the interaction region to see, for instance, if one
can probe arbitrarily short distances.
The full interpretation reveals subtle surprises.

The fixed-angle, fixed-genus calculation shows that the mean size
of the region probed by the string is of order
$$
   \vev{X}\sim \sqrt{s}/G
   .\eqn\?
$$
This shows that strings stretch at high energy (for fixed $G$),
but it also suggests
high orders processes in perturbation theory
probe arbitrarily short spacetime distances for a given energy~$\sqrt{s}$.
In fact this hope is dashed by large quantum fluctuations which render
the perturbation series divergent.
This divergence signals that the theory is unstable
at such short distances.
The series may be Borel resummed\refmark{\MO}, with the result
that the physically meaningful ``master trajectory'' has a size
that varies only logarithmically with energy, so we can crudely write
$$
   \vev{X}\sim {\cal O}(\lpl)
   ,\eqn\?
$$
while the amplitude behaves as
${\cal A}_{\rm resum}\sim \exp(-\sqrt{s})$, just on the edge of what
would be needed to restore locality.

What does this tell us about gravity and about geometry?

If short distances are unobservable, are they irrelevant?
Not necessarily.
We have seen that at Planck energies there is a limit to  extracting
short-distance information with a readily available spacetime
interpretation.
Fundamental though this regime is, we have to look elsewhere
to get a handle on how geometry is revealed in string
interactions over Planck distances.
We do see that it is crucial,
when the probes in the theory are themselves extended,
to ask only questions that can be answered with test strings.

\Chapter{3.}{Tidal Forces on Strings}

\noindent
Photons fall freely along geodesics.
Strings do not.
If string theory indeed describes how gravity acts on elementary particles
at the quantum level, we should explore the consequences
of this simple fact.

The first statement is an immediate consequence of the principle of
equivalence.  It leads to testable and tested observations:
bending of light by the sun, gravitational redshift, cosmological
redshift, etc.
To a structureless point particle in free fall, no gravitational
forces are present.

For extended objects, life is quite different.
Even in free fall they sense the presence of a gravitational
field through tidal forces.  These are, to be sure, very weak
in the case of a body whose characteristic size~$\ell$ is much much less
than the scale~$\cal R$  over which the field varies.
Yet small though these forces may be, they are never negligible compared
to zero and they may give the leading signature for the existence
of internal structure.

Consider deflection of light by the sun.
By the equivalence principle, we know that light of all frequencies
is deflected by an identical angle.
Indeed, the equation of motion for a classical, massless particle is
$$
   {\d ^2x^\mu \over \d \tau^2} + \Gamma^\mu_{\nu\lambda}
   {\d x^\nu\over \d\tau} {\d x^\lambda\over \d\tau}
   =0
   ,\eqn\Geodesic
$$
where $\tau$ is an affine parameter and $\Gamma^\mu_{\nu\lambda}$
is the Christoffel symbol for the Schwarzschild metric appropriate
to the region outside the sun.
A change in energy can be
absorbed in a change of affine parameter, so that the spacetime path is
independent of energy.
The is true regardless of the field equations satisfied by the metric.

Now for a classical string $x^\mu(\tau)$ becomes $X^\mu(\tau,\sigma)$,
Eq.~$\Geodesic$ generalizes to
$$
   \square X^\mu + \Gamma^\mu_{\nu\lambda}(X(\tau,\sigma)) \partial_a X^\nu
   \partial^a X^\lambda = 0
   ,\eqn\Stringeq
$$
so there are {\it always\/} forces, even in a frame
falling freely with the center of mass.
This is as expected for an extended body.
In Eq.~$\Geodesic$ one can find coordinates in which the connection
vanishes along the trajectory and the equation of motion is
$\xddot^\mu(\tau)=0$, but in Eq.~$\Stringeq$, the connection
cannot be made to vanish along the whole world sheet.
The equations of motion are non-linear (except for such special
cases as plane-wave backgrounds \refmark{\horowitz}), mixing all string modes.
The center-of-mass energy does not scale out and the geodesics
are no longer frequency-independent.
For photons in string theory, then,
one would expect to see colors bent by different amounts, and in
general a host of other effects, such as a frequency-dependence to
spectral redshifts.

Though these effects are extraordinarily small (typically suppressed
by $\ell/{\cal R} \sim M_{\rm Planck}/M_{\rm star}$),
they represent the leading deviation from
classical gravity due to finite structure at the Planck scale.
They are worth exploring.
Moreover, one sees that the string scale naturally limits the
principle of equivalence because there are no ``arbitrarily small,''
local, inertial frames.

To see stringy corrections to general relativity,
one usually considers how Einstein's field equations,
$$
   R_{\mu\nu}=0
   ,\eqn\?
$$
are modified by a series of correction terms, $\alpha'R R+\ldots$,
in powers of $\alpha'$, which arise from imposing conformal invariance
on the two-dimensional sigma model.
The new metric which solves the full string equations naturally
changes the point-particle trajectories of Eq.~$\Geodesic$
by terms of order $\alpha'$, but it does not change the fact
that the new trajectories are still independent of energy.
The changes themselves are not only small; they are
model-dependent and they are difficult even in principle
to distinguish from possible non-string modifications to Einstein's
equations.

The string-corrections predicted by Eq.~$\Stringeq$, on the other hand,
offer the hope of observing effects which have zero background in
classical gravity.
We concentrate, therefore,
on how the geodesics depend on the finite string size,
rather than on how the metric and trajectories
differ from those of general relativity.
One even obtains a good qualitative view
by ignoring all corrections to the metric equations,
and this is how we shall illustrate these issues.
It almost goes without saying that the type of string theory,
the compactification, the presence of supersymmetry can change only
the quantitative details, not the overall picture.
For clarity we discuss the massless vector states of the open bosonic
string but the heterotic string could be just as easily used.

Of course string theory predicts many departures from
classical gravity as well as from local point-particle quantum field theory.
If we could execute in practice any thought experiment at all,
we might look directly to the physics of high energies and temperatures.
But more subtle, indirect low-energy effects of the underlying theory
are perhaps more likely to be seen first and may well be essential in giving
direction to further exploration.


The motion of a quantized test string in a background gravitational field,
described by the metric~$G_{\mu\nu}$ is given by the
Polyakov path integral
$$
   \int{\cal D}X{\cal D}g\, \exp\left({i\over 2\pi\alpha'\hbar}
   \int\d^2\sigma\,\sqrt{g}g^{ab}\partial_a X^\mu
   \partial_b X^\nu G_{\mu\nu}(X) \right)
   .\eqn\pathint
$$
We work at string tree level and
assume for the moment that the metric $G_{\mu\nu}$ is
a conformally invariant background,
such that the $\beta$-function vanishes to all orders;
we comment briefly below on other possible couplings in this action.
Latin indices represent world-sheet coordinates,
$(\sigma^0, \sigma^1) = (\tau, \sigma)$
and Greek indices denote target space-time coordinates.
Both spacetime and worldsheet have Minkowski signature.

Initial and final states are imposed as boundary conditions
on the trajectories included in the sum or
by inserting vertex operators.
Alternatively, one may derive the Bogoliubov transformation
between the in and out states.
The coefficients which give the transition amplitude can be
extracted as usual from the time evolution equations
for the quantized modes.
This has been discussed in detail for the non-zero modes of the
string in Refs.~[\devega,\horowitz,\shockwave].
We will only discuss here the zero-mode dynamics,
which is given by the solving for the zero-mode propagation equation,
subject to the normal-ordered Virasoro constraints.

{}From Eq.~$\pathint$, the
equation of motion for a string $X(\tau,\sigma)$,
generalizing~$\Geodesic$ is
$$
   \square X^\mu + \Gamma^\mu_{\nu\lambda}(X) \partial_a X^\nu
   \partial^a X^\lambda = 0
   .\eqn\Stringeq
$$
For a pointlike string trajectory, \ie, when $X$ is independent
of $\sigma$, Eq.~$\Stringeq$ reduces to Eq.~$\Geodesic$.

The best way to see what is going on is to consider the motion of a small
string.  By small, we always mean small compared to the typical
scale of gravitational variations.  One expects the string's
center of mass to follow at least approximately
the same motion as a test particle, so let us expand
the string trajectory in the following way\refmark{\devega}:
$$
   X^\mu(\tau, \sigma) = x^\mu(\tau) \, + \, \eta^\mu(\tau,\sigma)
   \, + \, \xi^\mu(\tau,\sigma)
   .\eqn\Expand
$$

The first term, $x^\mu(\tau)$, is geodesic
satisfying Eq.~$\Geodesic$.
The merit of this expansion is that the connection $\Gamma$ and its
derivatives are all evaluated along $x^\mu(\tau)$.

We define the second term, $\eta$,
to be the solution to the  linearized
equation of motion expanded about $x^\mu$:
$$
   {\D ^2\eta^\mu \over \D \tau^2}
   - {\partial^2 \eta^\mu \over \partial\sigma^2}
   + R^\mu_{\alpha\nu\beta} \xdot^\alpha\xdot^\beta\eta^\nu = 0
   ,\eqn\Etaeq
$$
where
$\D /\D\tau$ represents the usual covariant derivative along the
curve $x^\mu(\tau)$.
The meaning of $\eta$ is clear from this equation:
The first two terms generalize the flat-space harmonic motion of the string
to include parallel transport,
and the curvature term represents the usual force of geodesic deviation
on an extended object.

The remaining piece of the expansion, $\xi$, carries the interesting physics.
It satisfies
$$
   \eqalign{
   {\D ^2\xi^\mu \over \D \tau^2}
   & - {\partial^2 \xi^\mu \over \partial\sigma^2}
   + R^\mu_{\alpha\nu\beta} \xdot^\alpha\xdot^\beta\xi^\nu
   =
   \cr &
   - \Gamma^\mu_{\nu\lambda}\partial_a\eta^\nu\partial^a\eta^\lambda
   - {1\over 2}\left(\partial_\alpha\partial_\beta\Gamma^\mu_{\nu\lambda}
   \right) \xdot^\nu\xdot^\lambda\eta^\alpha\eta^\beta
   -\left(\partial_\alpha\Gamma^\mu_{\nu\lambda} \right)\xdot^\nu
   \etadot^\lambda\eta^\alpha + \cdots
   ,\cr}\eqn\Xieq
$$
where we have only shown terms up to second order in the perturbative
expansion around $x^\mu(\tau)$.
The equation is linear in $\xi^\mu$ but not homogeneous.
The left-hand side has the same form as Eq.~$\Etaeq$.
After solving Eq.~$\Etaeq$,
the right-hand side, quadratic in $\eta$, acts as a source term
that alters the center-of-mass zero-mode motion of the string.

Indeed, Fourier expand $\eta(\tau,\sigma)$,
$$
   \eta^\mu(\tau,\sigma) = \sum \eta^\mu_m(\tau) \,{\rm e}^{i m\sigma}
   ,\eqn\?
$$
and similarly for $\xi$.
The center of mass of the string is then follows
$$
   x^\mu(\tau) + \eta_0^\mu(\tau)  + \xi_0^\mu(\tau)
   .\eqn\?
$$
Modulo reparametrizations, it is $\eta_0+\xi_0$ that departs from
the point-particle path.

The modes of $\eta$ are decoupled:
$$
   \eqalign{
   & {\D ^2\eta_m^\mu \over \D \tau^2} + m^2 \eta_m^\mu
   + R^\mu_{\alpha\nu\beta} \xdot^\alpha\xdot^\beta\eta_m^\nu
   \cr & \quad =
   \etaddot_m^\mu + m^2 \eta_m^\mu
   + 2 \Gamma^\mu_{\nu\lambda} \xdot^\nu\etadot_m^\lambda
   + \left(\partial_\beta \Gamma^\mu_{\nu\lambda}\right)
   \xdot^\nu \xdot^\lambda \eta_m^\beta
   \cr & \quad = 0
   .}\eqn\?
$$
$\eta$ describes the
transverse string oscillator modes, so that $\eta_0=0$
and $\eta_m \cdot x=0$ in an asymptotic flat space region.
For a scattering interaction this initial condition on Eq.~$\Etaeq$ ensures
the $\eta_0(\tau)=0$ throughout for all $\tau$; the first-order excitations
of the oscillators do not affect the center of mass.
This is intuitively clear:
tidal forces stretch an object but they do not shift its center of mass,
to leading order.

On the other hand, the higher order term $\xi_0$
does couple to the oscillator non-zero modes of $\eta$,
$$
   \eqalign{
   & {\D ^2\xi_0^\mu \over \D \tau^2}
   + R^\mu_{\alpha\nu\beta} \xdot^\alpha\xdot^\beta\xi_0^\nu
   \cr & \quad =
   - \Gamma^\mu_{\nu\lambda}
   \left( \etadot_{-m}^\nu \etadot_m^\lambda
   + m^2 \eta_{-m}^\nu \eta_m^\lambda \right)
   - {1\over 2}\left(\partial_\alpha\partial_\beta\Gamma^\mu_{\nu\lambda}
   \right) \xdot^\nu\xdot^\lambda\eta_{-m}^\alpha\eta_m^\beta
   \cr & \qquad
   -\left(\partial_\alpha\Gamma^\mu_{\nu\lambda} \right)\xdot^\nu
   \etadot_{-m}^\lambda\eta_m^\alpha
   .}\eqn\?
$$
It is $\xi_0(\tau)$, therefore, which gives the leading deviation
between the path followed by a string and the path followed by a particle
in a given gravitational field.

It is easy to see now that whether we use an exact metric solution
or an approximate one, say tree-level in sigma-model perturbation
theory, the source terms for $\xi_0$ will persist.
The coefficients of the equations will just get small corrections.

What about other background fields?
Obviously a Brans-Dicke scalar field just changes the equations
determining the metric, not the string propagation.
A massive dilaton, as one might hope for from non-perturbative effects,
poses no threat, and even a strictly massless dilaton can be absorbed
into the metric by a spacetime conformal rescaling, although this
can only be done for one type of massless matter.


\OMIT{
The classical description used thus far is
naturally adequate for big strings with high
occupation number and bears similarities to the dynamics
of cosmic strings.
}
For first-quantized strings, these classical equations
characterize the dominant contribution to the path integral,
Eq.~$\pathint$,
but must of course be supplemented by the constraint equations
$$
   T_{ab}=0
   \eqn\?
$$
to select the physical states.
To treat low-lying states
at the level of these calculations is necessary to
impose the correct quantum, normal-ordered, Virasoro conditions
in the asymptotic initial region.
A massless `photon' state, for instance, is given by
exciting a single mode of the lowest oscillator level:
$$
   \zeta_\mu\,\alpha_{-1}^\mu|0;p\rangle
   ,\eqn\?
$$
where
$\alpha_{-1}$ is the standard mode creation operator,
$\zeta_\mu$ is the polarization vector and
$|0;p\rangle$ is the vacuum state with no oscillator excitations
and momentum~$p$.
To compute the transition coefficients, then, one needs
to know $\eta_{-1}(\tau)$ and $\xi_0(\tau)$.
\footnote{{\rm b}}{\eightrm\baselineskip=10pt
Note that the classical version of this state (for circular polarization)
$$
   (X^0,X^1,X^2,X^3) = (\ap E, \cos\tau\cos\sigma, \sin\tau\cos\sigma,
   \ap E)
$$
is forbidden by the classical constraint
$\dot X^2 + {X'}^2 = 0$,
which lacks the normal-ordering constant that renders the spin-one
state massless.
}

\Section{3.1}{String motion in a Schwarzschild metric}

\noindent
Now we illustrate these effects with an example:
we apply these equations to motion in a background Schwarzschild
metric,
$$
   \d s^2 = G_{\mu\nu}\d x^\mu\d x^\nu
   = \left(1-{2G M\over r}\right)\d t^2 -
   \left(1-{2G M\over r}\right)^{-1}\d r^2 - r^2 \d\Omega^2
   .\eqn\?
$$
We specialize to the case of four non-compact dimensions for
concreteness.  It is convenient to denote
$$
   a(r) \equiv 1 - 2G M/r
   ,\eqn\?
$$
and to work in units where the
Schwarzschild radius $R_S=2G M=1$.

\Section{3.2}{Radial infall}

\noindent
The simplest case is radial infall
of a massless photon from infinity.
For a point particle, the motion determined
from $\d s^2=0$ is
$$
   \eqalign{
   \xdot^0 = \ap E/ a(r), \qquad \rdot  = -\ap E
   .\cr}\eqn\?
$$
A string photon of energy~$E$ in the asymptotic
region~$r\to\infty$ has, in addition to this center of mass motion,
a single transverse oscillator excitation.
\OMIT{
, so that at large distance,
$$
   (X^0,X^1,X^2,X^3) \to (\ap E, \cos\tau\cos\sigma, \sin\tau\cos\sigma,
   \ap E)
   \eqn\?
$$
}
This determines the initial conditions on $\eta$, and from the
equation of motion (cf. Eq.~$\Etaeq$),
$$
   \square\eta^\theta + 2\Gamma^\theta_{r\theta}\rdot\etadot^\theta=0
   \eqn\?
$$
one finds the solution
$$
   \eqalign{
   \eta^\theta(\tau,\sigma) &= {\cos\tau\cos\sigma\over \ap E\tau}\tilde\eta
   , \cr
   \eta^\phi(\tau,\sigma) &= {\sin\tau\cos\sigma\over \ap E\tau}\tilde\eta
   ,\cr}\eqn\?
$$
where $\tilde\eta \propto \lpl$ is a constant.
Now we can substitute this into the equation of motion for the zero mode of
$\xi$ to obtain

$$
   \eqalign{
   &\xiddot^r_0 + \xidot^r_0\left(\ap E a'\over a\right) + \xidot^0_0(\ap E a)
   + \xi^r_0\left(\ap E a'\over a\right)^2
    = {\tilde\eta^2 r a(r)\over 2\tau^4}\ap E ,
   \cr
   &\xiddot^0_0 - \xidot^0_0\left(\ap E a'\over a\right) + \xidot^r_0(\ap E a)
   + \xi^0_0\left(\left({a'\over a}\right)^2 - {a''\over a} \right)^2
   {(\ap E)^2 \over a} = 0
   .\cr}\eqn\?
$$

For radial motion, the function $r(\tau)$ is known,
$r=-\ap E\tau$.
These equations become clearer if we
eliminate the parameter $\tau$ in favor of $r$, since then
the energy~$E$ scales out of the equations!
One can solve for the scaling function, $\xi_0(r)$, numerically.
It evolves smoothly from infinity, at least in the weak field region
where we trust the formalism.
Note further that $\tilde\eta\propto \lpl$, so $\xi_0(r)\propto\lpl$.

The result of the second-order tidal effect $\xi_0$ is to change
the energy of the photon.
A static observer at radius~$r$ sees the string-photon with energy
$$
   \eqalign{
   E' &= {1\over\pi\ap}{\d\over\d\tau}\int_0^\pi\d\sigma\,X^0(\tau,\sigma)
   \cr &
   = {E \over a(r)} + {1\over\ap}{\d\xi_0\over\d\tau}
   \cr &
   = E \left( {1\over 1-R_s/r} - {\d\xi^0_0\over\d r}\right)
   .\cr}
   \eqn\?
$$
So the gravitational redshift is modified by a new term, whose size
is of order $\lpl^2/R_S^2$ --- as expected and extremely small.

The importance of this result is that the shift~$\Delta E$
is proportional to the energy~$E$, as is the usual redshift.
Had this not been the case, we would have predicted that the redshifts
of spectral lines from stars would depend on their frequency,
in contrast with classical gravity and giving --- in principle ---
a small deviation to signal the presence of string effects and quantum
gravity.

Is this true in general?
It would be disappointing if tidal effects all entered with the same
energy dependence as point particles.
Luckily this is not the case.
The energy-dependence for radial infall is a consequence of symmetry,
and it no longer holds for scattering.
The heuristic picture of classical spinning strings is helpful here.
In flat space, two strings with different energies
moving with the speed of light
differ in their rate of rotation.
High energy strings rotate more slowly when seen by an inertial
observer, as required by Lorentz invariance.
Now for two strings which fall radially inward and rotate
transversely, the tidal forces of Eq.~$\Stringeq$ are also
transverse.  Since they are independent of the string's orientation
in the transverse plane,
they are therefore independent of the rate
of rotation and hence independent of the energy.

\Section{3.3}{Deflection at large distance}

\noindent
To see what happens in a more general situation, consider
deflection of a massless string-photon at large impact parameter.

In classical gravity, any metric theory predicts that the deflection
of light by the sun, or other massive body, depends only on the
impact parameter, $b=L/E$, where $L$ is the angular momentum.
Light of all frequencies is deflected by the same angle.
The precise angle depends on the field equations of the metric,
but this fact does not; it depends on the principle of
equivalence and on the pointlike nature of the test particle.
Does light in string theory get dispersed as well as getting
deflected?

Physically, two limiting cases may be distinguished.
At extremely high energies, the string rotates very slowly as it passes
a star or black hole and tidal forces
stretch the string along a fixed direction.
At very low energies, the string rotates rapidly and is subjected to
alternate pulling and pushing.  The motion can then be
treated adiabatically.  The net tidal forces do not cancel; but
because the string undergoes many oscillations before the background
field changes appreciably, the net effect is independent of
just how fast the string is rotating.  That is, at low energies
the tidal effect on the deflection angle is energy-independent.
This is how the point particle limit is recovered.

Now the first case, a string at large enough energy that it rotates
slowly during the time of the interaction, requires alas
that the string energy exceed the rest mass of the black hole,
rendering the test-string treatment obviously inappropriate, and
suggesting where the calculation matches on to the physics of
high-energy string scattering.

In the low-energy case, we just argued physically that
the leading correction to the deflection angle is energy independent,
so we must go to higher adiabatic order and study the equations of motion
derived above.

Again, the center of mass starting trajectory for the expansion is
given (up to a quadrature) by
$$
   \eqalign{
   \rdot &= \pm \ap E\left( 1 - {b^2 a(r)\over r^2}\right)^{1/2} ,
   \cr
   \xdot^0 &= \ap E / a(r),
   \cr
   \phidot &= \ap E {b \over r^2}
   .\cr}
   \eqn\?
$$

Rather than use the string decomposition of Eq.~$\Expand$,
it proves slightly simpler to redefine
$\eta^\mu + \xi^\mu \to A^\mu + B^\mu$, where $A^\mu(\tau,\sigma)$
is defined to be the parallel transported oscillation of the string,
so that
$$
   {\D ^2 A^\mu \over \D \tau^2}
   + {\partial^2  A^\mu \over \partial\sigma^2}
   =0
   ,\eqn\Aeq
$$
and the geodesic deviation term present in Eq.~$\Etaeq$ now will appear
on the right-hand side
as a source term for $B^\mu$.

The resulting equations are still coupled and complicated.
The equation for $B^\phi$ is
$$
   \eqalign{
   \Bddot^\phi & + {2\rdot\over r}\Bdot^\phi + {2\phidot\over r}\Bdot^r
   - {2\rdot\phidot \over r^2} B_0^r
   \cr &
   = {\tilde A^2\over 4r^2} \Biggl[
   \cos^2\tau \left( \phidot^2\sin 2\phi + {2\phidot\rdot\over r} \cos^2\phi
   \right)
   + \cos 2\tau\sin 2\phi
   \cr & \qquad\qquad
   + \sin 2\tau\left(\phidot\cos 2\phi -
   {\rdot\over 2r}\sin 2\phi\right) \Biggr]
   .\cr}
   \eqn\?
$$

The terms on the right-hand side with $\cos\tau$ and $\sin\tau$
vary extremely rapidly; if one averages over several string oscillations,
only the coefficient of the $\cos^2\tau$ survives.
Since it is quadratic in $\tau$ derivatives of the coordinates,
the resulting equation can be rescaled, as in the radial infall
case, to show that the net change in the deflection angle
$B_0^\phi$ is independent of energy.

This is again smaller than terms we have ignored, so we must
work harder to obtain the leading energy-dependent contribution
to $B_0^\phi$.  This comes in next order of the adiabatic
expansion, giving a source term proportional to $(\ap E)^2$,
and hence an energy-dependent contribution to the scattering angle.
This is as expected from the discussion above.
We therefore expand the source term in time derivatives (and $B$ in
powers of $\ap$) to separate the overall cumulative shift in the source
term from the rapid oscillations.
These equation are lengthy and, in any event, no easier to solve
analytically.
Having identified the leading energy-dependent term in the scattering
angle, going further requires a more detailed analysis of the equations
than we will undertake here.  It is straightforward to check numerically
how it behaves and to verify that the source term is indeed physical.

\Chapter{4.}{Discussion}

\noindent
String theory predicts departures from general relativity not only because
conformal invariance modifies the field equations but because
strings themselves are extended objects.
This gives a characteristic signature of stringy effects at energies high
and low.

It should be noted that a related type of classical behavior was
considered long ago.  In Refs.~[\papapetrou] it was noted
that the equations of motion for a spinning particle differ from
Eq.~$\Geodesic$ and that this would lead, for instance, to an energy
dependent scattering angle in the light-deflection problem, if only the
spin orientation were perpendicular to the plane of motion --- which it is not.

Of course even if one had a small black hole and effects were much bigger,
there are numerous other effects that would have
to be computed to isolate the tidal physics; some of these are very large.
Loop effects of course make real photons much much larger
that the Planck length, even before considering wave optics effects.
To find any measurable effects it would be desirable to find
systems on which the tidal effects are cumulative.

\Chapter{}{Acknowledgements}

\noindent
It is a pleasure to thank R.~Brandenberger, F.~Low, A.~Tseytlin, and
A.~Vilenken for helpful conversations.
Research supported in part by
the U.S.~Department of Energy under contract DE-AC02-76-ER03130.

\ninerm
\def\it{\nineit}
\def\bf{}
\def\fourteenrm{}

\refout
\end

\def\refout{\par\penalty-400\vskip\chapterskip
   \spacecheck\referenceminspace
   \ifreferenceopen \Closeout\referencewrite \referenceopenfalse \fi
   \line{\fourteenrm\hfil REFERENCES\hfil}\vskip\headskip
   \input \jobname.refs
   }

\def\refout{
   \ifreferenceopen \Closeout\referencewrite \referenceopenfalse \fi
   \input \jobname.refs
   }